\documentstyle[12pt,ssi]{article}
\input epsf.tex
\begin{document}
\title{Quantum Gravity at the Planck Length}
\author{Joseph Polchinski\thanks{Supported by NSF Grants PHY94-07194 and
PHY97-22022.}\\ 
Institute for Theoretical Physics\\
University of California\\
Santa Barbara, CA 93106-4030}
\maketitle
\begin{abstract}I describe our understanding of physics near
the Planck length, in particular the great progress of the last four years
in string theory.  Lectures presented at the 1998 SLAC Summer Institute.
\end{abstract}
\section{Introduction}
For obvious reasons, the SLAC Summer Institute is usually concerned
with the three particle interactions.  It is very appropriate, though,
that the subject of the 1998 SSI is gravity, because the next step in
understanding the weak, strong, and electromagnetic interactions will
probably require the inclusion of gravity as well.
There are many reasons for making this statement, but I will focus on
two, one based on supersymmetry and one based on the unification of
the couplings.

What is supersymmetry?
I will answer this in more detail later, but for now let me give two
short answers:
\begin{itemize}
\item[A.] A lot of new particles.
\item[B.] A new {\it spacetime} symmetry.
\end{itemize}
Answer A is the pragmatic one for a particle experimentalist or
phenomenologist.  In answer B, I am distinguishing internal symmetries
like flavor and electric charge, which act on the fields at each point
of spacetime, from symmetries like Lorentz invariance that move the
fields from one point to another.  Supersymmetry is of the second
type.  If the widely anticipated discovery of supersymmetry actually
takes place in the next few years, it not only means a lot more
particles to discover.  It also will be the first new spacetime
symmetry since the discovery of relativity, bringing the structure of
the particle interactions closer to that of gravity; in a sense,
supersymmetry is a partial unification of particle physics and gravity.

The unification of the couplings is depicted in figure~1.
\begin{figure}
\begin{center}
\leavevmode
\epsfbox{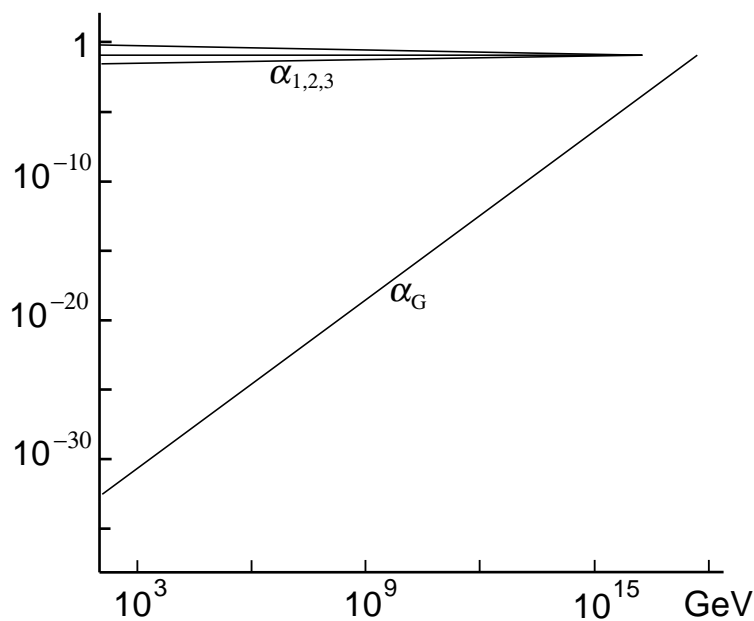}
\caption{The three gauge couplings and the dimensionless gravitational
coupling as functions of the energy.  Here $\alpha_{\rm G} = 
G_{\rm N} E^2 / 8\pi$.}
\end{center}
\end{figure}
This is
usually drawn with a rather different vertical scale.  Here the scale
is compressed so that the three gauge couplings can hardly be
distinguished, but this makes room for the fourth coupling, the
gravitational coupling.  Newton's constant is dimensionful, so what is
actually drawn is the dimensionless coupling $G_{\rm N} E^2$ with $E$
the energy scale and $\hbar = c=1$.  This dimensionless
gravitational coupling depends strongly on energy, in contrast to the
slow running of the gauge couplings.

It is well-known that the three
gauge couplings unify to good accuracy (in supersymmetric theories) at
an energy around $2 \times 10^{16}$ GeV.  Note however that the fourth
coupling does not miss by much, a factor of 20 or 30 in energy scale.  This
is another way of saying that the grand unification scale is near the Planck
scale.  In fact, the Planck scale $M_{\rm P} = 2 \times 10^{19}$ GeV is
deceptively high because of various factors like $4\pi$ that must be
included.  Figure~1 suggests that the grand unification of the three
gauge interactions will actually be a very grand unification including
gravity as well.  The failure of the four couplings to meet exactly
could be due to any of several small effects, which I will discuss
briefly later.\footnote{I will also discuss briefly the idea of low
energy string theory, in which figure~1 is drastically changed.}

Figure~1 also shows why the phenomenologies of the gauge interactions
and gravity are so different: at accessible energies the coupling
strengths are very different.  For the same reason, the energy scale
where the couplings meet is far removed from experiment. 
Nevertheless, we believe that we can deduce much of what happens at
this scale, and this is the subject of my lectures.  At the end I will
briefly discuss experimental signatures, and Michael Peskin and Nima
Arkani-Hamed will discuss some of these in more detail.

In section~2 I discuss the idea that spacetime has more than four
dimensions: first why this is not such a radical idea, and then why it is
actually a good idea.  In section~3 I review string theory as it stood a
few years ago: the motivations from the short distance problem of gravity,
from earlier unifying ideas, and from the search for new mathematical
ideas, as well as the main problem, vacuum selection.  In sections~4 I
introduce the idea of duality, including weak--strong and
electric--magnetic.  I explain how supersymmetry gives information about
strongly coupled systems.  I then describe the consequences for string
theory, including string duality, the eleventh dimension, D-branes, and
M-theory.  In section~5 I develop an alternative theory of quantum gravity,
only to find that `all roads lead to string theory.'  In section~6 I
explain how the new methods have solved some of the puzzles of black
hole quantum mechanics.  This in turn leads to the Maldacena dualities,
which give detailed new information about supersymmetric gauge field
theories.  In section~7 I discuss some of the ways that the new ideas might
affect particle physics, through the unification of the couplings and the
possibility of low energy string theory and large new dimensions. 
In section~8 I summarize and present the outlook. 

\section{Beyond Four Dimensions}

Gravity is the dynamics of spacetime.  It is very likely that at
lengths near the Planck scale ($L_{\rm P} = 10^{-33}$ cm) it becomes
evident that spacetime has more than the four dimensions that are
visible to us.  That is, spacetime is as shown in figure~2a, with four
large dimensions (including time) and some additional number of small
and highly curved spatial dimensions.
\begin{figure}
\begin{center}
\leavevmode
\epsfbox{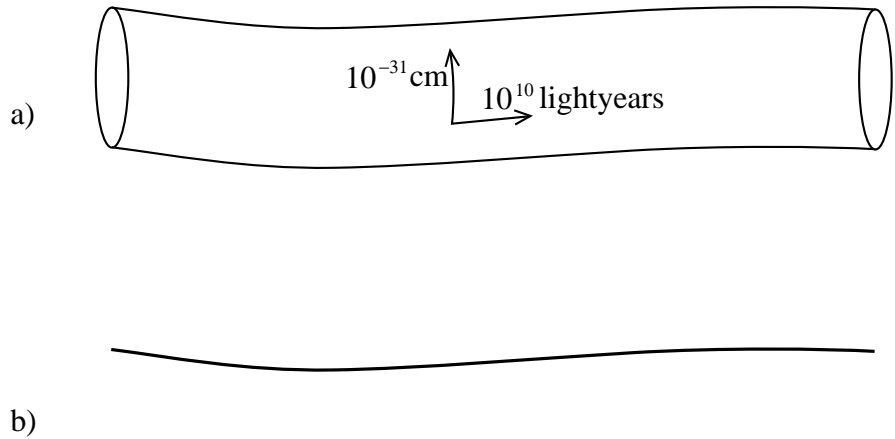}
\caption{a) A spacetime with one large dimension and one small one.  We assume
here that the small dimensions are nearly Planck sized; the possibility of
larger dimensions will be considered later.
 b) The same spacetime as seen by a low energy observer.}
\end{center}
\end{figure}
A physicist who probes this
spacetime with wavelengths long compared to the size of the small
dimensions sees only the large ones, as in figure~2b.
I will first give two reasons why this is a natural possibility to
consider, and then explain why it is a good idea.

The first argument is cosmological.  The universe is
expanding, so the dimensions that we see were once smaller and highly
curved.  It may have been that initially there were more than four
small dimensions, and that only the four that are evident to us began
to expand.  That is, we know of no reason that that the initial
expansion had to be isotropic.

The second argument is based on symmetry breaking.  Most of the
symmetry in nature is spontaneously broken or otherwise hidden from
us.  For example, of the $SU(3) \times SU(2) \times U(1)$ gauge
symmetries, only a $U(1)$ is visible.
Similarly the flavor symmetry is partly broken, as are the symmetries
in many condensed matter systems.  This symmetry breaking is part of
what makes physics so rich: if all of the symmetry of the underlying
theory were unbroken, it would be much easier to figure out what that
theory is!

Suppose that this same symmetry breaking principle holds for the
spacetime symmetries.  The visible spacetime symmetry is $SO(3,1)$, the
Lorentz invariance of special relativity consisting of the boosts and
rotations.  A larger symmetry would be $SO(d,1)$ for $d > 3$, the
Lorentz invariance of $d+1$ spacetime dimensions.  Figure~2 shows how
this symmetry would be broken by the geometry of spacetime.

So extra dimensions are cosmologically plausible, and are a natural
extension of the familiar phenomenon of spontaneous symmetry breaking.
In addition, they may be responsible for some of the physics that we
see in nature.  To see why this is so, consider first the following cartoon
version of grand unification.  The traceless $3\times3$ and $2\times2$
matrices for the strong and weak gauge interactions fit into a $5\times5$ 
matrix, with room for an extra $U(1)$ down the diagonal:
\begin{equation}
\left[
\begin{array}{ccc}
&\vrule&\\[-5pt]
3\times3&\vrule&\hspace{-4pt}X,Y\hspace{-2pt}\\[-5pt]
&\vrule&\\
\hline
&\vrule&\\[-10pt]X,Y&\vrule&\hspace{-9pt}2\times2\hspace{-4pt}\\[-10pt]
&\vrule&
\end{array}
\right]
\end{equation}
Now let us try to do something similar, but for gravity and
electromagnetism.  Gravity is described by a metric $g_{\mu\nu}$,
which is a $4\times4$ matrix, and electromagnetism by a 4-vector
$A_\mu$.  These fit into a $5\times5$ matrix:
\begin{equation}
\left[
\begin{array}{ccc}
&\vrule&\\[-5pt]
\hspace{5pt}g_{\mu\nu}\hspace{5pt}&\vrule&\hspace{-7pt}A_\mu\hspace{-3pt}
\\[-5pt]
&\vrule&\\
\hline&\vrule&\\[-15pt]
A_\nu&\vrule&\hspace{-9pt}\phi\hspace{-4pt}\\[-15pt]&\vrule&
\end{array}
\right] \label{kk}
\end{equation}
In fact, if one takes Einstein's equations in five dimensions, and
writes them out in terms of the components~(\ref{kk}), they become
Einstein's equations for the four-dimensional metric $g_{\mu\nu}$ plus
Maxwell's equation for the vector potential $A_\mu$.  This elegant
unification of gravity and electromagnetism is known as {\it
Kaluza--Klein theory}. 

If one looks at the Dirac equation in the higher-dimensional
space, one finds a possible explanation for another of the striking
patterns in nature, the existence of quark and lepton generations. 
That is, a single spinor field in the higher-dimensional space generally
reduces to several four-dimensional spinor fields, with repeated copies
of the same gauge quantum numbers.

Unification is accompanied by new physics. 
In the case of grand unification this includes the $X$ and $Y$ bosons,
which mediate proton decay.  In Kaluza--Klein theory it includes the
dilaton $\phi$, which is the last element in the matrix~(\ref{kk}).  I
will discuss the dilaton further later, but for now let me note that it
is likely not to have observable effects.  Of course, in Kaluza--Klein
theory there is more new physics: the extra dimension(s)!

Finally, let me consider the threshold behavior as one passes from
figure~2b to figure~2a.  At energies greater than the inverse size of
the small dimensions, one can excite particles moving in those
directions.  The states are quantized because of the finite size, and
each state of motion looks, from the lower-dimensional point of view,
like a different kind of particle.  Thus the signature of passing such
a threshold is a whole tower of new particles, with a
spectrum characteristic of the shape of the extra dimensions.

\section{String Theory}

\subsection{The UV Problem}

To motivate string theory, I will start with the UV problem of quantum
gravity.  A very similar problem arose in the early days of the weak
interaction.  The original Fermi theory was based on an interaction of four
fermionic fields at a spacetime point as depicted in figure~3a.
\begin{figure}
\begin{center}
\leavevmode
\epsfbox{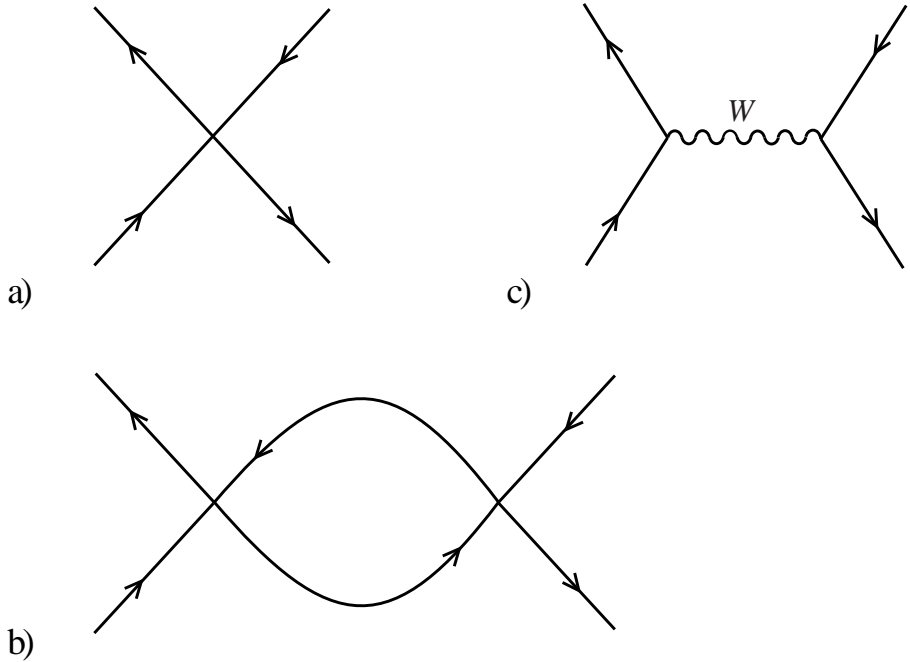}
\caption{a) A leptonic weak interaction in four-fermi theory.
b) A divergent second order amplitude.
c) The weak interaction in Weinberg-Salam theory.  At short distance
the contact interaction is resolved into the exchange of a $W$ boson.}
\end{center}
\end{figure}
The Fermi coupling constant $G_{\rm F}$ has units of
length-squared, or inverse energy-squared.  In a process with a
characteristic energy $E$ the effective dimensionless coupling is
then $G_{\rm F} E^2$.  It follows that at sufficiently high energy the
coupling becomes arbitrarily strong, and this also implies divergences
in the perturbation theory.  The second order weak amplitude of figure~3b is
dimensionally of the form 
\begin{equation} G_{\rm F}^2 \int^\infty E' dE', 
\end{equation} 
where $E'$ is the energy of the virtual state in
the second order process, and this diverges at high energy.  In
position space the divergence comes when the two weak interactions
occur at the same spacetime point (high energy = short distance).  The
divergences become worse at each higher order of perturbation theory
and so cannot be controlled even with renormalization.

Such a divergence suggests that the theory one is
working with is only valid up to some energy scale, beyond which new physics
appears.  The new physics should have the effect of smearing out the
interaction in spacetime and so softening the high energy behavior.  One might
imagine that this could be done in many ways, but in fact it is quite
difficult to do without spoiling Lorentz invariance or causality; this is
because Lorentz invariance requires that if the interaction is spread out in
space it is also spread out in time.  The solution to the short-distance
problem of the weak interaction is not quite unique, but combined with two of
the broad features of the weak interaction --- its
$V-A$ structure and its universal coupling to different quarks and
leptons --- a unique solution emerges.  This is depicted in figure~3c,
where the four-fermi interaction is resolved into the
exchange of a vector boson.  Moreover, this vector boson must be of a
very specific kind, coming from a spontaneously broken gauge
invariance.  And indeed, this is the way that nature works.\footnote{It
could also have been that the divergences are an artifact of
perturbation theory but do not appear in the exact amplitudes.  This is
a logical possibility, a `nontrivial UV fixed point.'  Although
possible, it seems unlikely, and it is not what happens in the case
of the weak interaction.}

For gravity the discussion is much the same.  The gravitational interaction
is depicted in figure~4a.
\begin{figure}
\begin{center}
\leavevmode
\epsfbox{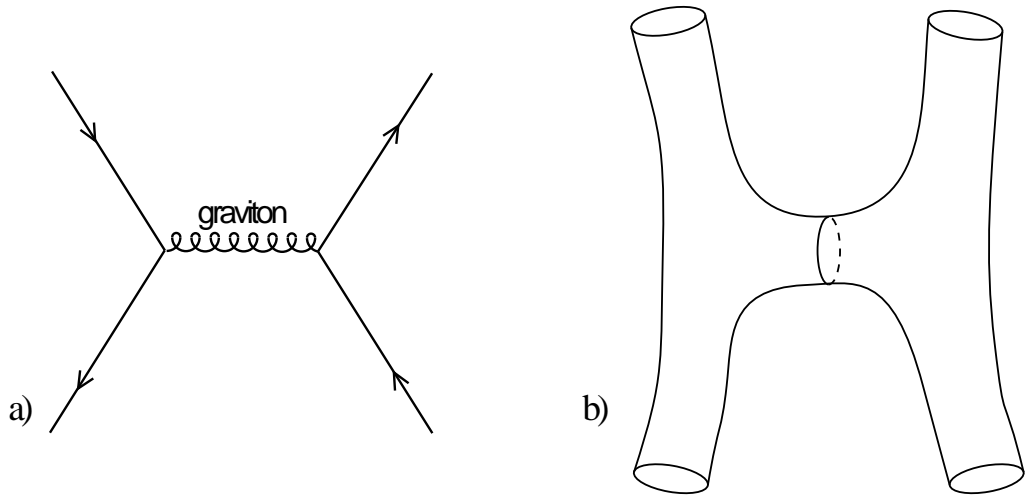}
\caption{a) Exchange of a graviton between two elementary particles.
b) The same interaction in string theory.  The amplitude is given by
the sum over histories, over all embeddings of the string
world-sheet in spacetime.  The world-sheet is smooth:
there is no distinguished point at which the interaction occurs (the
cross section on the intermediate line is only for illustration).}
\end{center}
\end{figure}
As we have already noted in
discussing figure~1, the
gravitational  coupling $G_{\rm N}$ has units of length-squared and so
the dimensionless coupling is $G_{\rm N} E^2$.  This grows large at
high energy and gives again a nonrenormalizable perturbation
theory.\footnote{Note that the bad gravitational interaction of
figure~4a is the same graph as the smeared-out weak interaction of
figure~3c.  However, its high energy behavior is worse because gravity
couples to energy rather than charge.}  Again the natural suspicion is
that new short-distance physics smears out the interaction, and again
there is only one known way to do this.  It
involves a bigger step than in the case of the weak interaction: it
requires that at the Planck length the graviton and other particles
turn out to be not points but one-dimensional objects, loops of
`string,' figure~5a.
\begin{figure}
\begin{center}
\leavevmode
\epsfbox{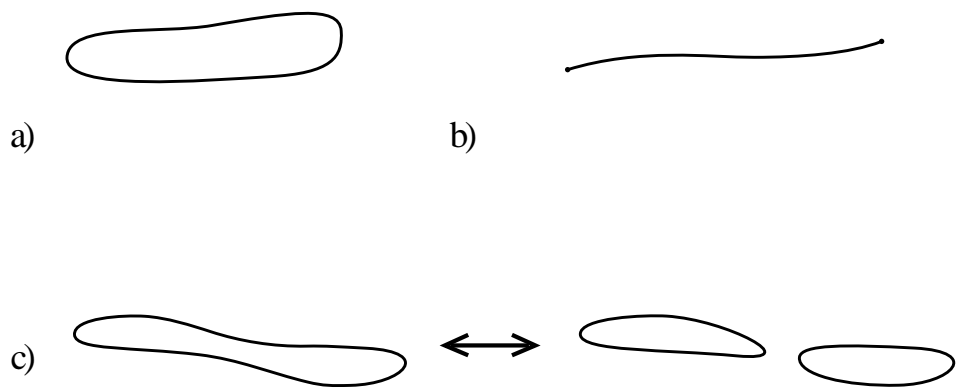}
\caption{a) A closed loop of string.  b) An open string, which appears in
some theories.  c) The basic splitting--joining interaction.}
\end{center}
\end{figure}
Their spacetime histories are then two-dimensional
surfaces as shown in figure~4b.

At first sight this is an odd idea.  It is not obvious why it should work
and not other possibilities.  It may simply be that we have not been
imaginative enough, but because UV problems are so hard to solve we should
consider carefully this one solution that we have found.  And in this case
the idea becomes increasingly attractive as we consider it.

\subsection{All Roads Lead to String Theory}

The basic idea is that the string has
different states with the properties of different particles.  Its
internal vibrations are quantized, and depending on which oscillators are
excited it can look like a scalar, a gauge boson, a graviton, or a fermion.
Thus the full Standard Model plus gravity can be obtained from this one
building block.  The basic string interaction is as in figure~5c, one string
splitting in two or the reverse.  This one interaction, depending on the
states of the strings involved, can look like any of the interactions in
nature: gauge, gravitational, Yukawa.

A promising fact is that string theory is unique: we have known for
some time that there are only a small number of string theories, and now have
learned that these are actually all the same.  (For now, this does not lead
to predictive power because the theory has many vacuum states, with
different physics.)

Further, string theory dovetails very nicely with previous ideas for
extending the Standard Model.  First, string theory automatically
incorporates supersymmetry: it turns out that in order for the theory to
be consistent the strings must move in a `superspace' which has `fermionic'
dimensions in addition to the ordinary ones.  Second, the spacetime symmetry
of string theory is $SO(9,1)$, meaning that the strings move in ten
dimensions.  As I have already explained, this is a likely way to explain
some of the features of nature, and it is incorporated in string theory.
Third, string theory can incorporate ordinary grand unification: some of the
simplest string vacua have the same gauge groups and matter that one finds
in unifying the Standard Model.

From another point of view, if one searches for higher symmetries or new
mathematical structures that might be useful in physics, one again finds many
connections to string theory. 
It is worthwhile to note that these three kinds of
motivation --- solving the divergence problem, explaining the broad patterns
in the Standard Model, and the connection with mathematics, were
also present in the weak interaction.  Weinberg emphasized the
divergence problem as I have done.  Salam was more guided by the
idea that non-Abelian gauge theory was a beautiful mathematical structure that
should be incorporated in physics.  Experiment gave no direct
indication that the weak interaction was anything but the pointlike
interaction of figure~3a, and no direct clue as to the new physics
that smears it out, just as today it gives no direct indication of
what lies beyond the Standard Model.  But it did show certain broad
patterns --- universality and the $V-A$ structure --- that were telltale
signs that the weak interaction is due to exchange of a gauge boson. 
It appears that nature is kind to us, in providing many trails to a
correct theory.

\subsection{Vacuum Selection and Dynamics}

So how do we go from explaining broad patterns to making precise
predictions?  The main problem is that string theory has many approximately
stable vacua, corresponding to different shapes and sizes for the rolled-up
dimensions.  The physics that we see depends on which of these vacua we are
in.  Thus we need to understand the dynamics of the theory in great detail,
so as to determine which vacua are truly stable, and how cosmology selects
one among the stable vacua.

Until recently our understanding of string theory was based entirely on
perturbation theory, the analog of the Feynman graph expansion, describing
small numbers of strings interacting weakly.  However, we know from quantum
field theory that there are many important dynamical effects that arise when
we have large numbers of degrees of freedom and/or strong couplings.  Some of
these effects, such as confinement, the Higgs mechanism, and dynamical
symmetry breaking, play an essential role in the Standard Model.  If one did
not know about them, one could not understand how the Standard Model
Hamiltonian actually gives rise to the physics that we see.

String theory is seemingly much more complicated than field theory, and so
undoubtedly has new dynamical effects of its own.
I am sure that all the experimentalists would like to know, ``How do I
falsify string theory?  How do I make it go away and not come back?''
Well, you can't.  Not yet.  To understand why, remember that in the '50s
Wolfgang Pauli thought that he had falsified Yang--Mills theory, because it
seems to predict long range forces not seen in nature.  The field
equations for the weak and strong forces are closely parallel to those for
electromagnetism, and so apparently of infinite range.  It is the
dynamical effects, symmetry breaking and confinement, that make these short
range forces.  Just as one couldn't falsify Yang--Mills theory in the '50s,
one cannot falsify string theory today.  In particular, because we cannot
reach the analog of the parton regime where the stringy physics is directly
visible, the physics that we see is filtered through a great deal of
complicated dynamics.

There is a deeper problem as well.  The Feynman graph expansion does not
converge, in field theory or string theory.  Thus it does not define the
theory at finite nonzero coupling.  One needs more, the analog of the path
integral and renormalization group of field theory.

Happily, since 1994 we have many new methods for understanding both field
theories and string theory at strong coupling.  These have led to steady
progress on the questions that we need to answer, and to many new results and
many surprises.  This progress is the subject of the rest of my lectures.

\section{Duality in Field and String Theory}

\subsection{Dualities}

One important idea in the recent developments is {\it duality.}  
This refers to the equivalence between seemingly distinct physical systems.
One starts with different Hamiltonians, and even with different fields, but
when after solving the theory one finds that the spectra and the transition
amplitudes are identical.  Often this occurs because a quantum system has
more than one classical limit, so that one gets back to the same quantum
theory by `quantizing' either classical theory.

This phenomenon is common in quantum field theories in two spacetime
dimensions.  The duality of the Sine-Gordon and Thirring models is one
example; the high-temperature--low-temperature duality of the Ising model is
another. The great surprise of the recent developments is that it is
also common in quantum field theories in four dimensions, and in string
theory.

A particularly important phenomenon is {\it weak--strong duality.}  I have
emphasized that perturbation theory does not converge.  It gives the
asymptotics as the coupling $g$ goes to zero, but it misses important physics
at finite coupling, and at large coupling it becomes more and more useless.
In some cases, though, when $g$ becomes very large there is a simple
alternate description, a weakly coupled dual theory with $g' = 1/g$.
In one sense, as $g \to \infty$ the quantum fluctuations of the original
fields become very large (non-Gaussian), but one can find a dual set of
fields which become more and more classical.

Another important idea is {\it electric--magnetic duality.}  A striking
feature of Maxwell's equations is the symmetry of the left-hand side under
${\bf E} \to {\bf B}$ and ${\bf B} \to -{\bf E}$.  This
symmetry suggests that there should be magnetic as well as electric
charges.  This idea became more interesting with Dirac's discovery of the
quantization condition
\begin{equation}
q_{\rm e} q_{\rm m} = 2\pi n \hbar\ , \label{dirq}
\end{equation}
which relates the quantization of the electric charge (its equal magnitude
for protons and electrons) to the existence of magnetic monopoles.
A further key step was the discovery by 't Hooft and Polyakov that grand
unified theories predict magnetic monopoles.  These monopoles are
solitons, smooth classical field configurations.  Thus they look rather
different from the electric charges, which are the basic quanta: the
latter are light, pointlike, and weakly coupled while monopoles are
heavy, `fuzzy,' and (as a consequence of the Dirac quantization) strongly
coupled.

In 1977 Montonen and Olive proposed that in certain supersymmetric
unified theories the situation at strong coupling would be reversed: the
electric objects would be big, heavy, and strongly coupled and the magnetic
objects small, light and weakly coupled.  The symmetry of the sourceless
Maxwell's equations would then be extended to the interacting theory, with
an inversion of the coupling constant.  Thus electric--magnetic duality
would be a special case of weak--strong duality, with the magnetically
charged fields being the dual variables for the strongly coupled theory.

The evidence for this conjecture was circumstantial: no one could actually
find the dual magnetic variables.  For this reason the reaction to this
conjecture was skeptical for many years.  In fact the evidence remains
circumstantial, but in recent years it has become so much stronger that the
existence of this duality is in little doubt.

\subsection{Supersymmetry and Strong Coupling}

The key that makes it possible to discuss the strongly coupled theory is
{\it supersymmetry.}  One way to think about supersymmetry is in terms of
extra dimensions --- but unlike the dimensions that we see, and unlike the
small dimensions discussed earlier, these dimensions are `fermionic.'  
In other words, the coordinates for ordinary dimensions are real numbers
and so commute with each other: they are `bosonic;' the fermionic
coordinates instead satisfy
\begin{equation}
\theta_i \theta_j = - \theta_j \theta_i \ .
\end{equation}
For $i = j$ this implies that $\theta_i^2 = 0$, so in some sense these
dimensions have zero size.  This may sound rather mysterious but in
practice the effect is the same as having just the bosonic dimensions but
with an extra symmetry that relates the masses and couplings of fermions
to those of bosons.

To understand how supersymmetry gives new information about strong coupling,
let us recall the distinction between {\it symmetry} and {\it dynamics.} 
Symmetry tells us that some quantities (masses or amplitudes) vanish, and
others are equal to one another.  To actually determine the values of the
masses or amplitudes is a dynamical question.  In fact, supersymmetry
gives some information that one would normally consider dynamical.
To see this, let us consider in quantum theory the Hamiltonian operator
$H$, the charge operator $G$ associated with an ordinary symmetry like
electric charge or baryon number, and the operator $Q$ associated with a
supersymmetry.  The statement that $G$ is a symmetry means that it
commutes with the Hamiltonian,
\begin{equation}
[H, G] = 0\ .
\end{equation}
For supersymmetry one has the same,
\begin{equation}
[H, Q] = 0\ , \label{susy1}
\end{equation}
but there is an additional relation
\begin{equation}
Q^2 = H + G\ , \label{susy2}
\end{equation}
in which the Hamiltonian and ordinary symmetries appear on the right.
There are usually several $G$s and several $Q$s, so that
there should be additional indices and constants in these equations, but
this schematic form is enough to explain the point.  It is this second
equation that gives the extra information.  To see one example of this,
consider a state $|\psi\rangle$ having the special property that it is
neutral under supersymmetry:
\begin{equation}
Q |\psi\rangle = 0\ .
\end{equation}
To be precise, since we have said that there are usually several $Q$s, we
are interested in states that are neutral under {\it at least one} $Q$
but usually not all of them.  These are known as {\it BPS
(Bogomolnyi--Prasad--Sommerfield) states}.  Now take the expectation
value of the second relation~(\ref{susy2}) in this state:  
\begin{equation}
\langle \psi | Q^2 |\psi\rangle = \langle \psi | H |\psi\rangle + \langle
\psi | G |\psi\rangle \ .
\end{equation}
The left side vanishes by the BPS property, while the two terms on the
right are the energy $E$ of the state $|\psi\rangle$ and its charge $q$
under the operator $G$.  Thus
\begin{equation}
E = - q\ , \label{bpsen}
\end{equation}
and so the energy of the state is determined in terms of its charge.  But
the energy is a dynamical quantity: even in quantum mechanics we must
solve Schr\"odinger's equation to obtain it.  Here, it is determined
entirely by symmetry information.  (There is a constant of proportionality
missing in~(\ref{bpsen}), because we omitted it from~(\ref{susy2}) for
simplicity, but it is determined by the symmetry.)

Since the calculation of $E$ uses only symmetry information, it does not
depend on any coupling being weak: it an exact property of the theory. 
Thus we know something about the spectrum at strong coupling.  Actually,
this argument only gives the allowed values of $E$, not the ones that
actually appear in the spectrum.  The latter requires an extra step: we
first calculate the spectrum of BPS states at weak coupling, and then
adiabatically continue the spectrum: the BPS property enables us to follow
the spectrum to strong coupling.

The BPS states are only a small part of the spectrum, but by using this
and similar types of information from supersymmetry, together with
general properties of quantum systems, one can usually recognize a
distinctive pattern in the strongly coupled theory and so deduce the dual
theory.  Actually, this argument was already made by Montonen and Olive in
1977, but only in 1994, after this kind of reasoning was applied in a
systematic way in many examples starting with Seiberg, did it become clear
that it works and that electric--magnetic duality is a real property of
supersymmetric gauge theories.

\subsection{String Duality, D-Branes, M-Theory}

Thus far the discussion of duality has focussed on quantum field theory,
but the same ideas apply to string theory.  Prior to 1994 there were
various conjectures about duality in string theory, but after the
developments described above, Hull, Townsend, and Witten considered the
issue in a systematic way.  They found that for each strongly coupled string
theory (with enough supersymmetry) there was a unique candidate for a
weakly coupled dual.  These conjectures fit together in an
intricate and consistent way as dimensions are compactified, and evidence
for them rapidly mounted.  Thus weak--strong duality seems to be a general
property in string theory.

Weak--strong duality in field theory interchanged the pointlike quanta of
the original fields with smooth solitons constructed from those fields. 
In string theory, the duality mixes up various kinds of object: the basic
quanta (which are now strings), smooth solitons, black holes (which are
like solitons, but with horizons and singularities), and new stringy
objects known as {\it D-branes.}

The D-branes play a major role, so I will describe them in more detail.
In string theory strings usually move freely.  However, some string
theories also predict localized objects, sort of like defects in a
crystal, where strings can break open and their endpoints get stuck. 
These are known as D-branes, short for Dirichlet (a kind of boundary
condition --- see Jackson) membranes.  Depicted in figure~6, they can be
points (D0-branes), curves (D1-branes), sheets (D2-branes), or
higher-dimensional objects.  They are dynamical objects --- they can move,
and bend --- and their properties, at weak coupling, can be determined with
the same machinery used elsewhere in string theory.
\begin{figure}
\begin{center}
\leavevmode
\epsfbox{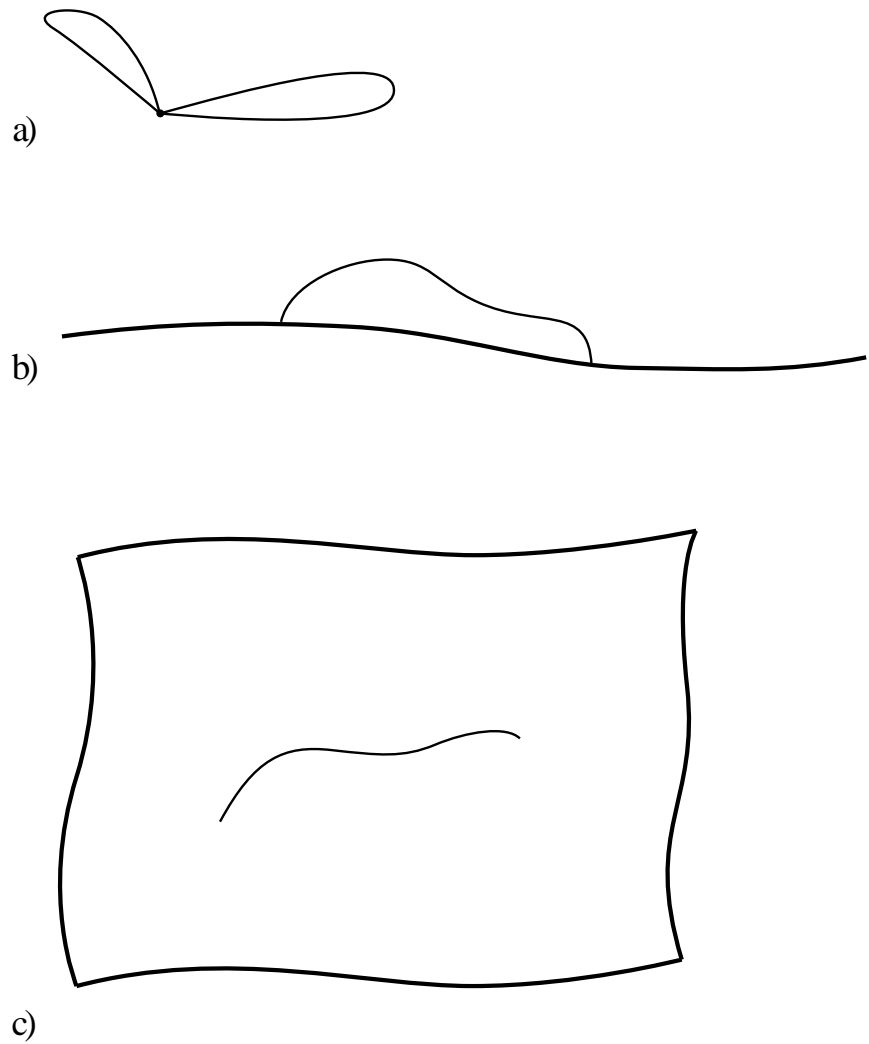}
\caption{a) A D0-brane with two attached strings.  b) A D1-brane (bold)
with attached string.  c) A D2-brane with attached string.}
\end{center}
\end{figure}

Even before string duality it was found that one could make D-branes
starting with just ordinary strings (for string theorists, I am talking
about
$T$-duality).  Now we know that they are needed to fill out the duality
multiplets.  They have many interesting properties.  One is that they are
smaller than strings; one cannot really see this pictorially, because it
includes the quantum fluctuations, but it follows from calculations of
the relevant form factors.  Since we are used to thinking that smaller means
more fundamental, this is intriguing, and we will return to it.

Returning to string duality, figure~7 gives a schematic picture of what
was learned in 1995.  Before that time there were five known string
theories.  These differed primarily in the way that supersymmetry acts on
the string, and the type I theory also in that it includes open strings.
\begin{figure}
\begin{center}
\leavevmode
\epsfbox{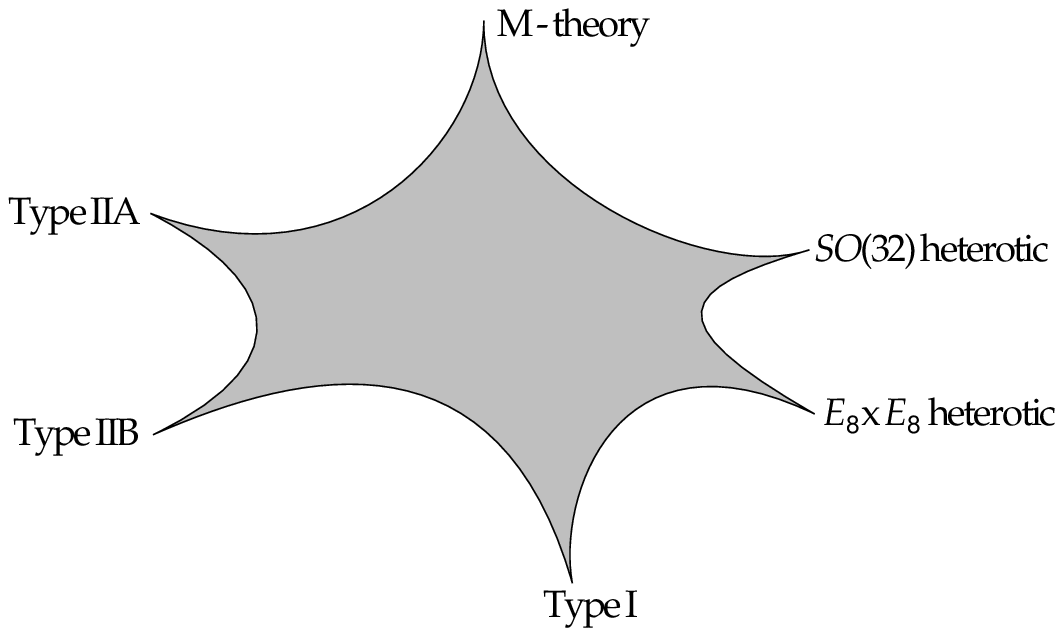}
\caption{The five string theories, and M-theory, as limits of a single
theory.}
\end{center}
\end{figure}
We now know that starting with any one of these theories and going to
strong coupling, we can reach any of the others.  Again, the idea is that
one follows the BPS states and recognizes distinctive patterns in the
limits.  The parameter space in the figure can be thought of as two
coupling constants, or as the radii of two compact dimensions.

In figure~7 there is a sixth limit, labeled {\it M-theory.}  We have
emphasized that the underlying spacetime symmetry of string theory is
$SO(9,1)$.  However, the M-theory point in the figure is in fact a point
of $SO(10,1)$ symmetry: the spacetime symmetry of string theory is larger
than had been suspected.  The extra piece is badly spontaneously broken,
at weak coupling, and not visible in the perturbation theory, but it is a
property of the exact theory.  It is interesting that $SO(10,1)$ is
known to be the largest spacetime symmetry compatible with supersymmetry.

Another way to describe this is that in the M-theory limit the theory
lives in eleven spacetime dimensions: a new dimension has appeared.  This
is one of the surprising discoveries of the past few years.  How does one
discover a new dimension?  It is worthwhile explaining this in some more
detail.  The D0-brane mass is related to the characteristic string mass
scale $m_{\rm s}$ and the dimensionless string coupling $g_{\rm s}$, by
\begin{equation}
m_{\rm D0} = \frac{m_{\rm s}}{g_{\rm s}}\ .
\end{equation}
When $g_{\rm s}$ is small this is heavier than the string scale, but when 
$g_{\rm s}$ is large it is lighter.  Further, the D0-brane is a BPS state
and so this result is exact.  If one considers now a state with $N$
D0-branes, the mass is bounded below by $N m_{\rm D0}$, an in fact this
bound is saturated: there is a BPS bound state with
\begin{equation}
m_{N\rm D0} = N\frac{m_{\rm s}}{g_{\rm s}}
\end{equation}
exactly.  Now observe that for $g_{\rm s}$ large, all of these masses
become small.  What can the physics be? In fact, this is the 
spectrum associated with passing a threshold where a new spacetime
dimension becomes visible.  The radius of this dimension is
\begin{equation}
R = \frac{g_{\rm s}}{m_{\rm s}}\ .
\end{equation}
That is, small $g_{\rm s}$ is small $R$ and large $g_{\rm s}$ is large
$R$.  In particular, perturbation theory in $g_{\rm s}$ is an expansion
around $R = 0$: this is why this dimension has always been invisible!

\section{An Alternative to String Theory?}

On Lance Dixon's tentative outline for my lectures, one of the items was
`Alternatives to String Theory.'  My first reaction was that this was
silly, there are no alternatives, but on reflection I realized that there
was an interesting alternative to discuss.
So let us try to construct a quantum theory of gravity based on a new
principle, not string theory.  We will fail, of course, but we will fail
in an interesting way.

Let us start as follows.  In quantum mechanics we have the usual
position-momentum uncertainty relation
\begin{equation}
\delta x \delta p \geq \hbar^2\ .
\end{equation}
Quantum gravity seems to imply a breakdown in spacetime at the Planck
length, so perhaps there is also a position-position uncertainty relation
\begin{equation}
\delta x \delta x \geq L_{\rm P}^2\ . \label{uncer}
\end{equation}
This has been discussed many times, and there are many ways that one might
try to implement it.  We will do this as follows.  Suppose that we have $N$
nonrelativistic particles.  In normal quantum mechanics the state would be
defined by $N$ position vectors
\begin{equation}
{\bf X}_i\ , \quad i = 1, \ldots, N\ .
\end{equation}
Let us instead make these into Hermitean matrices in the particle-number
index 
\begin{equation}
{\bf X}_{ij}\ , \quad i,j = 1, \ldots, N\ .
\end{equation}
It is not obvious what this means, but we will see that it
leads to an interesting result.  For the Hamiltonian we take
\begin{equation}
H = \frac{1}{2M} \sum_{m=1}^{D-1} \sum_{i,j = 1}^{N} (p_{ij}^m)^2
+ M'^5 \sum_{m,n=1}^{D-1} \sum_{i,j = 1}^{N} |[X^m,X^n]_{ij}|^2\ .
\label{math}
\end{equation}
The first term is just an ordinary nonrelativistic kinetic term, except
that we now have $N^2$ coordinate vectors rather than $N$ so there is a
momentum for each, and we sum the squares of all of them.  The indices $m$
and $n$ run over the $D-1$ spatial directions, and $M$ and $M'$ are large
masses, of order the Planck scale.  The potential term is chosen as
follows.  We want to recover ordinary quantum mechanics at low energy.
The potential is the sums of the squares of all of the components of all
of the commutators of the matrices ${\bf X}_{ij}$, with a large
coefficient.  It is therefore large unless all of these matrices commute.
In states with energies below the Planck scale, the matrices will then
commute to good approximation, so we do not see the new
uncertainty~(\ref{uncer}) and we recover the usual quantum mechanics.
In particular, we can find a basis which diagonalizes all the commuting
$X^m_{ij}$.  Thus the effective coordinates are just the $N$ diagonal
elements $X^m_{ii}$ of each matrix in this basis, which is the right count
for $N$ particles in ordinary quantum mechanics: the $X^m_{ii}$ behave
like ordinary coordinates. 

The Hamiltonian~(\ref{math}) has interesting connections with other parts
of physics.  First, the commutator-squared term has the exact same structure
as the four-gluon interaction in Yang--Mills theory.  This is no accident,
as we will see later on.  Second, there is a close connection to
supersymmetry.  In supersymmetric quantum mechanics, one has operators
satisfying the algebra~(\ref{susy1},\ref{susy2}).  Again in general there
are several supersymmetry charges, and the number $\cal N$ of these $Q$s 
is significant.  For small values of $\cal N$, like 1, 2 or 4, there are
many Hamiltonians with the symmetry.  As $\cal N$ increases the symmetry
becomes more constraining, and ${\cal N} = 16$ is the maximum number.
For ${\cal N} = 16$ there is only one invariant Hamiltonian, and it is
none other than our model~(\ref{math}).  To be precise, supersymmetry
requires that the particles have spin, that the Hamiltonian also has a
spin-dependent piece, and that the spacetime dimension $D$ be 10.
In fact, supersymmetry is necessary for this idea to work.  The vanishing
of the potential for commuting configurations was needed, but we only
considered the classical potential, not the quantum corrections.  The
latter vanish only if the theory is supersymmetric.

So this model has interesting connections, but let us return to the idea
that we want a theory of gravity.  The interactions among low energy
particles come about as follows.  We have argued that the potential forces
the ${\bf X}_{ij}$ to be diagonal: the off-diagonal pieces are very
massive.  Still, virtual off-diagonal excitations induce interactions
among the low-energy states.  In fact, the leading effect, from one loop
of the massive states, produces precisely the (super)gravity interaction
among the low energy particles.

So this simple idea seems to be working quite well, but we said that we
were going to fail in our attempt to find an alternative to string theory.
In fact we have failed because this is not an alternative: it {\it is}
string theory.  It is actually one piece of string theory, namely the
Hamiltonian describing the low energy dynamics of $N$ D0-branes.  This
illustrates the following principle: that all good ideas are part of
string theory.  That sounds arrogant, but with all the recent progress in
string theory, and a fuller understanding of the dualities and dynamical
possibilities, string theory has extended its reach into more areas of
mathematics and has absorbed previous ideas for unification
(including $D=11$ supergravity).

We have discussed this model not just to introduce this principle, but
because the model is important for a number of other reasons.  In fact, it
is conjectured that it is not just a piece of string theory, but is
actually a {\it complete} description.  The idea is that if we view
any state in string theory from a very highly boosted frame, it will be
described by the Hamiltonian~(\ref{math}) with $N$ large.  Particle
physicists are familiar with the idea that systems look different as one
boosts them: the parton distributions evolve.  The idea here is that the
D0-branes are the partons for string theory; in effect the string is a
necklace of partons.  This is the {\it matrix theory} idea of Banks,
Fischler, Shenker, and Susskind (based on earlier ideas of Thorn), and at
this point it seems very likely to be correct or at least a step in the
correct direction.

To put this in context, let us return to the illustration in
figure~7 of the space of string vacua, and to the point made earlier that
the perturbation theory does not define the theory for finite $g$.  In
fact, every indication is that the string description is useful only near
the five cusps of the figure in which the string coupling becomes weak. 
In the center of the parameter space, not only do we not know the
Hamiltonian but we do not know what degrees of freedom are supposed to
appear in it.  It is likely that they are not the one-dimensional objects
that one usually thinks of in string theory; is it more likely that they
are the coordinate matrices of the D-branes.

\section{Black Hole Quantum Mechanics}

\subsection{Black Hole Thermodynamics}

In the '70s it was found that there is a close analogy between the laws of
black hole mechanics and the laws of thermodynamics.  In particular, the
event horizon area (in Planck units) is like the entropy.  It is
nondecreasing in classical gravitational processes, and the sum of this
{\it Bekenstein--Hawking entropy} and the entropy of radiation is
nondecreasing when Hawking radiation is included.
For more than 20 years is has been a goal to find the statistical
mechanical picture from which this thermodynamics derives --- that is, to
count the quantum states of a black hole.  There have been suggestive
ideas over the years, but no systematic framework for addressing the
question.

I have described the new ideas we have for understanding strongly
interacting strings.  A black hole certainly has strong gravitational
interactions, so we might hope that the new tools would be useful here.
Pursuing this line of thought, Strominger and Vafa were able in early
1996 to count the quantum states of a black hole for the first time.  They
did this with the following thought experiment.  Start with a black hole
and imagine adiabatically reducing the gravitational coupling $G_{\rm N}$. 
At some point the gravitational binding becomes weak enough that the black
hole can no longer stay black, but must turn into ordinary matter.  A
complete theory of quantum gravity must predict what the final state will
look like.  The answer depends on what kind of black hole we begin with ---
in other words, what are its electric, magnetic, and other charges (the
no-hair theorem says that this is all that identifies a black hole).
For the state counting we want to take a supersymmetric black hole, one
that corresponds to a BPS state in the quantum theory.  For the simplest
such black holes, the charges that they carry determine that at weak
coupling they will turn into a gas of weakly coupled D-branes, as depicted
in figure~8.
\begin{figure}
\begin{center}
\leavevmode
\epsfbox{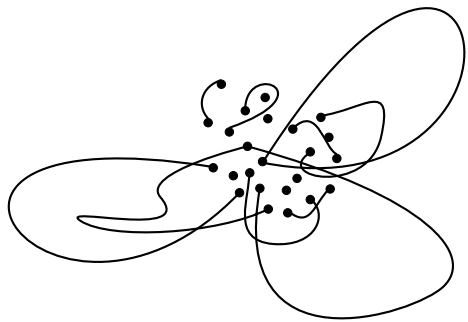}
\caption{A state of many D0-branes, with attached
strings.}
\end{center}
\end{figure}
For these we know the Hamiltonian, so we can count the states and continue
back to strong coupling where the system is a black hole.  Indeed, the
answer is found to agree precisely with the Bekenstein--Hawking entropy.
Our initial motivation was one problem of quantum gravity, the UV
divergences.  Now, many years later, string theory has solved another,
very different, long-standing problem in the subject.

This result led to much further study.  It was found that in addition to
the agreement of the entropy of BPS states with the Bekenstein--Hawking
entropy, agreement was also found between the general relativity
calculation and the D-brane calculation for the entropies of near-BPS
states, and for various dynamical quantities such as absorption and decay
amplitudes.  This goes beyond the adiabatic continuation argument used to
justify the entropy calculation, and in late 1997 these results were
understood as consequences of a new duality, the {\it Maldacena}
duality.  This states that, not only are the weakly coupled D-branes
the adiabatic continuation of the black hole, but that {\it the 
D-brane system at all couplings is dual (physically equivalent) to the
black hole.}  In effect, D-branes are the atoms from which certain black
holes are made, but for large black holes they are in a highly quantum
state while the dual gravitational field is in a very classical state.
A precise statement of the Maldacena duality requires a low energy
limit in the D-brane system, while on the gravitational side one takes the
limit of the geometry near the horizon.

\subsection{The Information Paradox}

This new duality has two important consequences.  The first is for another
of the nagging problems of quantum gravity, the black hole information
paradox. A black hole emits thermal Hawking radiation, and will eventually
decay completely.  The final state is independent of what went into the
black hole, and incoherent.  In other words, an initially pure state
evolves into a mixed state; this is inconsistent with the usual rules of
quantum mechanics.  Hawking argued that in quantum gravity the evolution
of states must be generalized in this way.

This has been a source of great controversy.  While most physicists would
be pleased to see quantum mechanics replaced by something less weird,
the particular modification proposed by Hawking simply makes it uglier,
and quite possibly inconsistent.  But twenty years of people trying to
find Hawking's `mistake,' to identify the mechanism that preserves the
purity of the quantum state, has only served to sharpen the paradox:
because the quantum correlations are lost behind the horizon,
either quantum mechanics is modified in Hawking's way, or the locality of
physics must break down in a way that is subtle enough not to infect most
of physics, yet act over long distances.

The duality conjecture above states that the black hole is equivalent to an
ordinary quantum system, so that the laws of quantum evolution are
unmodified.  However, to resolve fully the paradox one must identify the
associated nonlocality in the spacetime physics.  This is hard to do
because the local properties of spacetime are difficult to extract from
the highly quantum D-brane system: this is related to the {\it holographic
principle.}  This term refers to the property of a hologram, that the full
picture is contained in any one piece.  It also has the further connotation
that the quantum state of any system can be encoded in variables living on
the {\it boundary} of that system, an idea that is suggested by the
entropy--area connection of the black hole.  This is a key point where our
ideas are in still in flux.

\subsection{Black Holes and Gauge Theory}

Dualities between two systems give information in each direction: for each
system there are some things that can be calculated much more easily in
the dual description.  In the previous subsection we used the Maldacena
duality to make statements about black holes.  We can also use it in the
other direction, to calculate properties of the D-brane theory.

To take full advantage of this we must first make a generalization.  We
have said that D-branes can be points, strings, sheets, and so on: they
can be extended in $p$ directions, where here $p = 0, 1, 2$.  Thus we
refer to D$p$-branes.  The same is true of black holes: the usual ones are
local objects, but we can also have black strings --- strings with
event horizons --- and so on.  A black $p$-brane is extended in $p$
directions and has a black hole geometry in the orthogonal directions.
The full Maldacena duality is between
the low energy physics of D$p$-branes and strings in the near-horizon
geometry of a black $p$-brane.  Further, for $p \leq 3$ the low energy
physics of $N$ D$p$-branes is described by $U(N)$ Yang--Mills theory with
${\cal N} = 16$ supersymmetries.  That is, the gauge fields live on the
D-branes, so that they constitute a field theory in $p+1$ `spacetime'
dimensions, where here spacetime is just the world-volume of the brane.
For $p=0$, this is  the connection of matrix quantum mechanics to
Yang--Mills theory that we have already mentioned below~(\ref{math}).

The Maldacena duality then implies that various quantities in the gauge
theory can be calculated more easily in the dual black $p$-brane geometry.
This method is only useful for large $N$, because this is necessary to get
a black hole which is larger than string scale and so described by
ordinary general relativity.  Of course we have a particular interest in
gauge theories in $3+1$ dimensions, so let us focus on $p=3$.  The
Maldacena duality for $p=3$ partly solves an old problem in the strong
interaction.  In the mid-'70s 't Hooft observed that Yang--Mills theory
simplifies when the number of colors is large.  This simplification was
not enough to allow analytic calculation, but its form led 't Hooft to
conjecture a duality between large-$N$ gauge theory and some unknown
string theory.  The Maldacena duality is a precise realization of this
idea, for supersymmetric gauge theories.\footnote
{For $p=3$ the near-horizon geometry is the product of an anti-de Sitter
space and a sphere, while the supersymmetric gauge theory is conformally
invariant (a conformal field theory), so this is also known as the
{\it AdS--CFT correspondence.}}
For the strong
interaction we need of course to understand nonsupersymmetric gauge
theories.  One can obtain a rough picture of these from the Maldacena
duality, but a precise description seems still far off.  It is notable,
however, that string theory, which began as an attempt to describe the
strong interaction, have now returned to their roots, but only by means of
an excursion through black hole physics and other strange paths.

\subsection{Spacetime Topology Change}

This subsection is not directly related to black holes, but deals with
another exotic question in quantum gravity.  Gravity is due to the bending
of spacetime.  It is an old question, whether spacetime can not only bend
but break: does its topology as well as its geometry evolve in time?

Again, string theory provides the tools to answer this question.  The answer
is `yes' --- under certain controlled circumstances the geometry can evolve
as shown schematically in figure~9.
\begin{figure}
\begin{center}
\leavevmode
\epsfbox{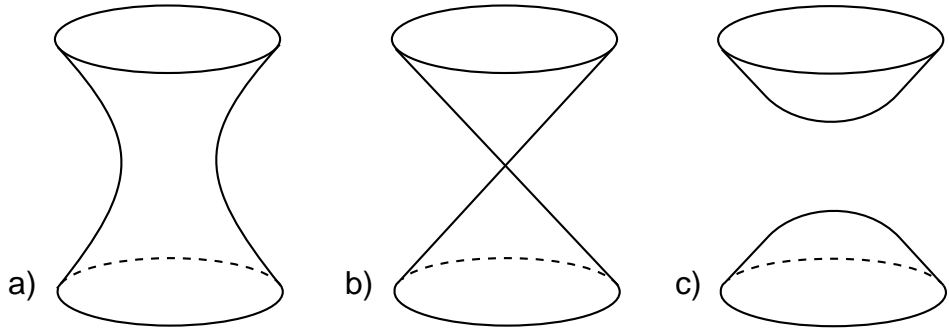}
\caption{An example of spacetime topology change.}
\end{center}
\end{figure}
It is interesting to focus on the case that the topology change is taking
place in the compactified dimensions, and to contrast the situation as
seen by the short-distance and long-distance observers of figures~2a
and~2b.  The short distance observer sees the actual process of figure~9. 
The long distance observer cannot see this.  Rather, this observer sees a
{\it phase transition.}  At the point where the topology changes, some
additional particles become massless and the symmetry breaking pattern
changes.  Thus the transition can be analyzed with the ordinary methods of
field theory; it is this that makes the quantitative analysis of the
topology change possible.

Incidentally, topology change has often been discussed in the context of
spacetime foam, the idea that the topology of spacetime is constantly
fluctuating at Planckian distance scales.  It is likely that the truth is
even more strange, as in matrix theory where spacetime becomes
`non-Abelian.'

\section{Unification and Large Dimensions}

My talks have been unapologetically theoretical.  The Planck length is far
removed from experiment, yet we believe we have a great deal of
understanding of the very exotic physics that lies there.  In this final
section I would like to discuss some ways in which the discoveries of the
last few years might affect the physics that we see.

Let me return to the unification of the couplings in figure~10a, and to the
failure of the gravitational coupling to meet the other three exactly.
\begin{figure}
\begin{center}
\leavevmode
\epsfbox{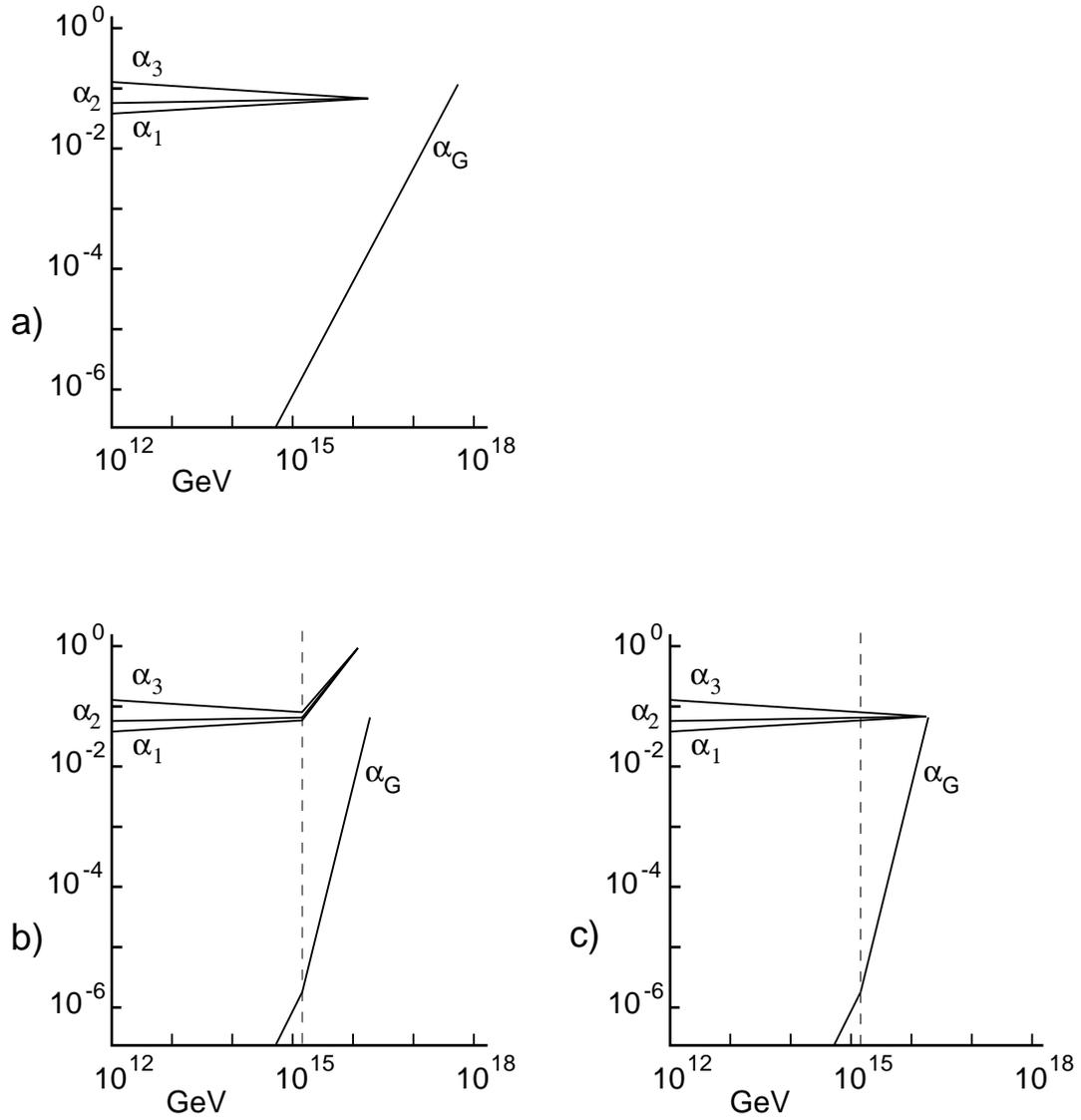}
\caption{a) Running of the three gauge couplings and the dimensionless
gravitational coupling with energy.  b) Effect of a fifth dimension
below the unification scale.  c) Effect of a fifth dimension of
Horava-Witten type.}
\end{center}
\end{figure}
There are many ideas to explain this.  There could be additional particles
at the weak, intermediate, or unified scales, which change the running
of the gauge couplings so as to raise the unification point.  Or it may be
that the gauge couplings actually do unify first, so that there is a
normal grand unified theory over a small range of scales before the
gravitational coupling unifies.
These ideas focus on changing the behavior of the gauge couplings.  Since
these already unify to good approximation, it would be simpler to change
the behavior of the gravitational coupling so that it meets the other
three at a lower scale --- to lower the Planck scale.  Unfortunately this
is not so easy.  The energy-dependence of the gravitational coupling is
just dimensional analysis, which is not so easy to change.\footnote{It was
asked whether the gravitational coupling has additional $\beta$-function
type running.  Although this could occur in principle, it does not do so
because of a combination of dimensional analysis and symmetry arguments.}

There is a way to change the dimensional analysis --- that is, to change
the dimension!  We have discussed the possibility that at some scale we
pass the threshold to a new dimension.  Suppose that this occurred below
the unification scale.  For both the gauge and the gravitational couplings
the units change, so that both turn upward as in figure~10b.  This does
not help; the couplings meet no sooner.

There is a more interesting possibility, which was first noticed in the
strong coupling limit of the $E_8 \times E_8$ heterotic string.  Of the
five string theories, this is the one whose weakly coupled physics looks
most promising for unification.  Its strong-coupling behavior, shown in
figure~11, is interesting.  A new dimension appears, but it is not simply a
circle.  Rather, it is bounded by two walls.  Moreover, all the gauge
fields and the particles that carry gauge charges move only in the walls,
\begin{figure}
\begin{center}
\leavevmode
\epsfbox{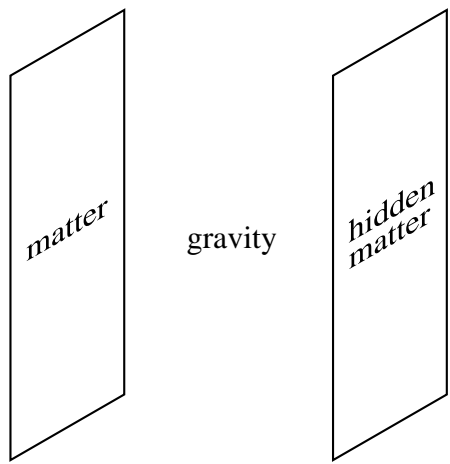}
\caption{A Horava--Witten spacetime.  The two planes represent $3+1$
dimensional walls, in which all the Standard Model particles live, while
gravity moves in the $4+1$ dimensional bulk between the walls.  In string
theory there are six additional dimensions, which could be much smaller
and are not shown.  The wall is then $9+1$ dimensional in all, and the
spacetime $10+1$ dimensional.}
\end{center}
\end{figure}
while gravity moves in the bulk.  Consider now the unification of the
couplings.  The dynamics of the gauge couplings, and their running,
remains as in $3+1$ dimensions; however, the gravitational coupling has a
kink at the threshold, so the net effect can be as in figure~10c.  If the
threshold is at the correct scale, the four couplings meet at a point.

As it stands this has no more predictive power than any of the other
proposed solutions.  There is one more unknown parameter, the new threshold
scale, and one more prediction.  However, it does illustrate that the new
understand of string theory will lead to some very new ideas about
the nature of unification.  Figure~11 is only one example of a much more
general idea now under study, that the Standard Model lives in a
brane and does not move in the full space of the compact dimensions, while
gravity does do so.

This new idea leads in turn to the possibility of radically changing the
scales of new physics in string theory.  To see this, imagine lowering the
threshold energy (the kink) in figure~10c; this also lowers the string
scale, which is where the gravitational coupling meets the other three. 
From a completely model-independent point of view, how low can we go?
The string scale must be at least a TeV, else we would already have seen
string physics.  The five-dimensional threshold must correspond to a
radius of no more than a millimeter, else Cavendish experiments would
already have shown the four-dimensional inverse square law turning into a
five-dimensional inverse cube.  Remarkably, it is difficult to improve on
these extreme model-independent bounds.  The large dimension in particular
might seem to imply a whole tower of new states at energies above
$10^{-4}$ eV, but these are very weakly coupled (gravitational strength)
and so would not be seen.
It may be that construction of a full model, with a sensible cosmology,
will raise these scales, but that they will still lie lower than we used
to imagine.

I had been somewhat skeptical about this idea, for a reason that is
evident in figure 10c.  If the threshold is lowered further, the
gravitational coupling meets the other three before they unify and one
loses the successful prediction of $\sin^2 \theta_{\rm w}$.  However, it is
wrong to pin so much on this one number; the correct prediction might come
out in the end in a more complicated way.  One should certainly explore the
many new possibilities that arise, to see what other consequences there are
and to broaden our perspective on the possible nature of unification.

\section{Outlook}

I will start with the more theoretical problems.

{\bf 1. The black hole information problem.}  It seems that the necessary
ingredients to solve this are at hand, and that we will soon assemble them
correctly. However, it has seemed this way before, and the clock on this
problem is at 22 years and counting.  Still, our understanding is clearly
deeper than it has ever been.

{\bf 2. The cosmological constant problem.}  In any quantum theory the vacuum
is a busy place, and should gravitate.  Why is the cosmological constant,
even if nonzero, so much smaller than particle or Planck energies?
This is another hard problem, not just in string theory but in any theory of
gravity.  It has resisted solution for a long time, and seems to require
radical new ideas.

The new ideas that I have described have not led to a solution, but they have
suggested new possibilities.  One important ingredient may be
supersymmetry.  Throughout the discussion of duality this plays a central
role in canceling quantum fluctuations, suggesting that it also does so in
the vacuum energy.  The problem is that supersymmetry is broken in nature;
we need a phase with some properties of the broken theory and some of the
unbroken.  We have learned about many new phases of string theory, but not
yet one with just the right properties.

Another ingredient may be
nonlocality.  The cosmological constant affects physics on cosmic scales
but is determined by dynamics at short distance: this suggests the need for
some nonlocal feedback mechanism.  Recall that the black hole information
problem also seems to need nonlocality; perhaps these are related.

{\bf 3. Precise predictions from string theory?}
Our understanding of string dynamics is much improved, but still very
insufficient for solving the vacuum selection/stability
problem, especially with nonsupersymmetric vacua.  It is hard to see how one
could begin to address this before solving the cosmological constant
problem, since this tells us that we are missing something important about
the vacuum.

An optimistic projection is that we soon solve the information problem,
that this gives us the needed idea to solve the cosmological constant
problem, and then we can address vacuum selection.  More likely,
we still are missing some key concepts.

{\bf 4. What is string theory?}
We are closer to a nonperturbative formulation than ever before: the things
that we have learned in the past few years have completely changed our
point of view.  It may be that again the ingredients are in place,
in that both matrix theory and the Maldacena duality give nonperturbative
definitions, and we simply need to extract the essence.

{\bf 5. Distinct signatures of string theory?}
Is there any distinctively stringy experimental signature? 
All of the new physics may lie far beyond accessible energies, but we might
be lucky instead.  I have discussed the possibility of low energy string
theory and large dimensions.  I am still inclined to expect the standard
picture to hold, but the new ideas are and will remain a serious
alternative.  Another possibility is a fifth force from the dilaton or
other moduli (scalars that are common in string theory).  These are
massless to first approximation, but quantum effects almost invariably
induce masses for all scalars.  The resulting mass is likely in the range
\begin{equation}
\frac{m_{\rm weak}^2}{m_{\rm P}} < m_{\rm scalar} < m_{\rm weak}\ .
\end{equation}
The lower limit is interesting for a fifth force, while the whole range is
interesting for dark matter.

The most interesting hope is for something unexpected, perhaps cosmological
and associated with the holographic principle, or perhaps a distinctive
form of supersymmetry breaking.

{\bf 6. Supersymmetry.} 
Supersymmetry has played a role throughout these talks.  In string theory it
is a symmetry at least at the Planck scale, but is broken somewhere between
the Planck and weak scales.  The main arguments for breaking at the weak
scale are independent of string theory: the Higgs hierarchy problem, the
unification of the couplings, the heavy top quark.  In addition, the
ubiquitous role that supersymmetry plays in suppressing quantum
fluctuations in our discussion of strongly coupled physics supports the
idea that it suppresses the quantum corrections to the Higgs mass.  The one
cautionary note is that the cosmological constant suggests a new phase of
supersymmetry, whose phenomenology at this point is completely unknown. 
Still, the discovery and precision study of supersymmetry remains the best
bet for testing all of these ideas.

In conclusion, the last few years have seen remarkable progress, and there
is a real prospect of answering difficult and long-standing problems in the
near future.

\section*{References}
\baselineskip=15pt

Two texts on string theory:
\begin{itemize}
\item M. B. Green, J. H.
Schwarz, and E. Witten, {\it Superstring Theory, Vols. 1 and~2}
(Cambridge University Press, Cambridge, 1987).
\item 
J. Polchinski, {\it String Theory, Vols. 1 and 2}
(Cambridge University Press, Cambridge, 1998).
\end{itemize}
An earlier version of these lectures, with extensive references:
\begin{itemize}
\item J. Polchinski,  ``String Duality,''  Rev. Mod. Phys. {\bf 68},
1245 (1996), hep-th/9607050.
\end{itemize}
The talk by Michael Peskin at the SSI Topical Conference gives more detail
on some of the subjects in my lectures.  Other popular accounts:
\begin{itemize}
\item G. P. Collins, ``Quantum Black Holes Are Tied to D-Branes and Strings,''
Physics Today {\bf 50}, 19 (March 1997).
\item 
E. Witten, ``Duality, Spacetime, and Quantum Mechanics,''
Physics Today {\bf 50}, 28 (May 1997).
\item 
B. G. Levi, ``Strings May Tie Quantum Gravity to Quantum Chromodynamics,''
Physics Today {\bf 51}, 20 (August 1998).
\end{itemize}
A summer school covering many of the developments up to 1996:
\begin{itemize}
\item
C. Efthimiou and B. Greene, editors, {\it Fields, Strings, and
Duality, TASI 1996} (Singapore: World Scientific, 1997). 
\end{itemize}
Lectures on matrix theory:
\begin{itemize}
\item
T. Banks, ``Matrix Theory,''
Nucl. Phys. Proc. Suppl. {\bf 67}, 180 (1998), hep-th/9710231.
\item
D. Bigatti and L. Susskind, ``Review of Matrix Theory,''
hep-th/9712072.
\end{itemize}
A discussion of the spacetime uncertainty principle:
\begin{itemize}
\item
M. Li and T. Yoneya, ``Short Distance Spacetime Structure and Black Holes
in String Theory,'' hep-th/9806240.
\end{itemize}
Lectures on developments in black hole quantum mechanics:
\begin{itemize}
\item
G. T. Horowitz, ``Quantum States of Black Holes,''
gr-qc/9704072.
\item
A. W. Peet,  
``The Bekenstein Formula and String Theory,''
Classical and Quantum Gravity {\bf 15}, 3291 (1998), 
hep-th/9712253.
\end{itemize}
A recent review of nonperturbative string theory:
\begin{itemize}
\item
A. Sen, ``An Introduction to Non-perturbative String Theory,''
hep-ph/9802051.
\end{itemize}
Two very recent colloquium-style presentations, including the Maldacena
conjecture:
\begin{itemize}
\item
A. Sen, ``Developments in Superstring Theory,'' hep-ph/9810356.
\item
J. Schwarz, ``Beyond Gauge Theories,'' hep-th/9807195. 
\end{itemize}
For more on low energy string theory and millimeter dimensions see the talk
by Nima Arkani-Hamed at the SSI Topical Conference, and references therein.

\end{document}